\documentclass[journal=jctcce]{achemso}

\usepackage{graphicx}
\usepackage{amsmath}
\usepackage{dcolumn}
\usepackage{bm}
\usepackage[dvipsnames]{xcolor}

\title{Reweighted Jarzynski sampling: Acceleration of rare events and free energy calculation with a bias potential learned from nonequilibrium work}

\author{Kristof M. Bal}
  \email{kristof.bal@uantwerpen.be}
  \affiliation{Department of Chemistry and NANOlab Center of Excellence, University of Antwerp, Universiteitsplein 1, 2610 Antwerp, Belgium}

\date{\today}

\begin{document}

\begin{tocentry}
\includegraphics{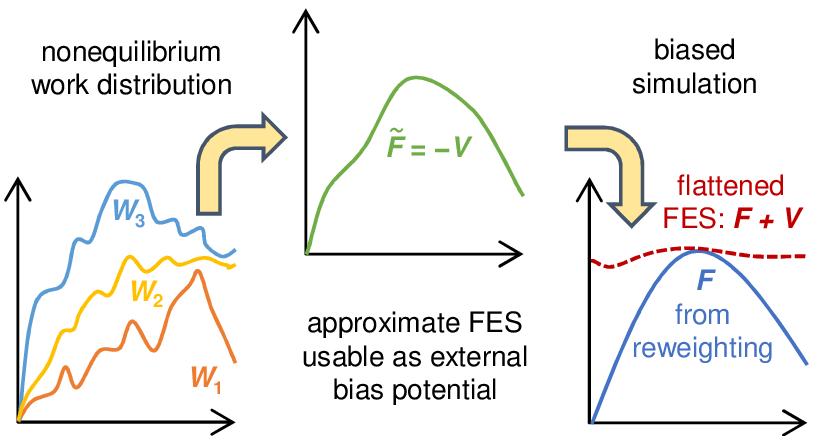}
\end{tocentry}

\begin{abstract}
We introduce a simple enhanced sampling approach for the calculation of free energy differences and barriers along a one-dimensional reaction coordinate.
First, a small number of short nonequilibrium simulations are carried out along the reaction coordinate, and the Jarzynski equality is used to learn an approximate free energy surface from the nonequilibrium work distribution. 
This free energy estimate is represented in a compact form as an artificial neural network and used as an external bias potential to accelerate rare events in a subsequent molecular dynamics simulation. 
The final free energy estimate is then obtained by reweighting the equilibrium probability distribution of the reaction coordinate sampled under the influence of the external bias. 
We apply our reweighted Jarzynski sampling recipe to four processes of varying scales and complexities---spanning chemical reaction in the gas phase, pair association in solution, and droplet nucleation in supersaturated vapor. 
In all cases, we find reweighted Jarzynski sampling to be a very efficient strategy, resulting in rapid convergence of the free energy to high precision.
\end{abstract}

\maketitle

\section{Introduction}

The free energy $F$ is a central thermodynamic quantity that governs the equilibrium properties and dynamics of a system.
In order to calculate free energy differences between states, one must be able to express the free energy as a low-dimensional function $F (\mathbf{s})$, in which $\mathbf{s}$ are collective variables that can properly distinguish configurations belonging to any of the states of interest.
To evaluate the free energy surface (FES) $F (\mathbf{s})$ using atomistic simulations, enough configurations for all values of $\mathbf{s}$ must be sampled.
This is, however, challenging if high free energy barriers are present on $F (\mathbf{s})$ and state-to-state transitions are rare events.

Several techniques have been developed in the last few decades to improve sampling and allow for an efficient reconstruction of $F (\mathbf{s})$.
One particular class is that of the so-called \emph{adaptive} sampling methods~\cite{Mezei1987,Darve2001,Laio2002,Valsson2014,Zhang2019}.
Typically, such a method uses an external bias potential or force.
This bias is refined iteratively in such a way that the system fully explores the whole collective variable space $\mathbf{s}$ of interest.
The best possible sampling can be achieved if the bias perfectly compensates $F (\mathbf{s})$ or its gradient, i.e., $V (\mathbf{s}) = -F (\mathbf{s})$.
Therefore, a converged bias in adaptive sampling methods is also a good estimator of the free energy.

Although the simultaneous optimization of sampling and reconstruction of the free energy is an attractive proposition, practical application of adaptive sampling methods can be difficult.
In essence, one must find a balance between fast exploration of configuration space (which requires rapid updates of the bias) and a smooth convergence of the free energy (slow updates of the bias), which may require extensive optimization of the simulation parameters~\cite{Dama2014}.
In addition, even if a method's bias is formally an estimator of the free energy, such as in metadynamics~\cite{Laio2002} or the adaptive biasing force algorithm~\cite{Darve2001}, it is not necessarily the most efficient.
Indeed, a free energy estimate from on-the-fly sampled forces or histogram tends to converge more quickly than the bias itself in many cases~\cite{Cuendet2014,Mones2016,Marinova2019}.

Convergence of the free energy requires frequent visits of all regions in configuration space~\cite{Cuendet2014}.
In this sense, it may be better to have a roughly reasonable bias potential as soon as possible: $V (\mathbf{s})$ need not be a perfect estimator of $F (\mathbf{s})$, as long as it is able to induce state-to-state transitions on an MD time scale.
If sampling is good, $F (\mathbf{s})$ can then be obtained from the average force or reweighted histogram.
Properly reweighting the action of a fluctuating adaptive bias potential is however challenging~\cite{Cuendet2014,Mones2016,Marinova2019}. 
In contrast, reweighting a trajectory obtained with a fixed, static bias potential---a concept dating back to umbrella sampling~\cite{Torrie1977}---is a much more robust approach, provided that the bias allows for sufficient sampling. 
It was therefore soon realized that an adaptive method can be used to generate $V (\mathbf{s})$, which then serves as a suitable static bias in a subsequent reweighting~\cite{Babin2006}.

Several recent sampling strategies have further explored the decoupling of the bias and the underlying FES~\cite{McCarty2016,Invernizzi2019,Debnath2020}.
A recurring concept in such methods is that an approximate FES $\widetilde{F}$ is constructed as an interpolation of local models of $F(\mathbf{s})$  for (well sampled) individual metastable states.
Generating an efficient sampling bias $V (\mathbf{s}) = -\widetilde{F}(\mathbf{s})$ then only entails optimizing a limited number of interpolation parameters.
Reweighting, finally, takes care of any remaining discrepancies between $F(\mathbf{s})$ and $\widetilde{F} (\mathbf{s})$.

In this work, we present a simple Ansatz for an efficient sampling bias that requires no adaptive optimization before or during the sampling trajectory.
In a two-state system, characterized by a reaction coordinate $\chi$, a bias $V (\chi)$ is learned from the work distribution of a small number of nonequilibrium molecular dynamics trajectories along $\chi$, using the Jarzynski equality~\cite{Jarzynski1997}. 
An artificial neural network (ANN) is used to represent a smooth bias potential in a flexible manner.
Although Jarzynski's free energy estimator is hard to converge reliably~\cite{Hummer2001,Gore2003,Park2004,Shirts2005,Lua2005,Oberhofer2005,Jarzynski2006}, we find that its second order cumulant expansion can efficiently generate a bias potential capable of driving transitions along $\chi$.
A simple reweighting of the biased trajectory suffices to reconstruct the unbiased free energy $F (\chi)$.
We demonstrate our reweighted Jarzynski sampling approach on four systems representing gas phase chemical reactions, processes in solution, and phase transitions.
For all systems, accurate free energy differences and barriers are obtained at a cost competitive to, or lower than, previous enhanced sampling studies.

\section{Theory}

\subsection{Jarzynski equality and free energy reconstruction}

Suppose a transformation of interest can be described by a single collective variable $\chi (\mathbf{R})$, in which $\mathbf{R}$ are the microscopic coordinates of the the system.
Let the system be equilibrated at a temperature $T$, in contact with a heat bath at $T$, and defined by an initial state $\chi = \chi_a$.
An external force applied to $\chi$ can then be used to drive the system to a new state $\chi_b$, delivering an amount of work $W$.
Jarzynski discovered the existence of a relation between the nonequilibrium work $W$ of such a process, and the free energy difference $\Delta F = F(\chi_b) - F(\chi_a)$ between the microstates $\chi_a$ and $\chi_b$~\cite{Jarzynski1997}:
\begin{equation}
  e^{-\beta \Delta F} = \langle e^{-\beta W} \rangle , \label{eq:Jarzynski}
\end{equation}
in which $\beta = (k_B T)^{-1}$ and $k_B$ the Boltzmann constant.
The average $\langle \cdots \rangle$ indicates that from a large number of \emph{nonequilibrium} experiments $\chi_a \rightarrow \chi_b$, the \emph{equilibrium} free energy difference $\Delta F$ can be recovered.

A practical difficulty of applying Eq.~\eqref{eq:Jarzynski} for the calculation of free energy differences, is the poor convergence of its exponential average~\cite{Jarzynski2006}.
A typical realization of the nonequilibrium experiment will require $W > \Delta F$ and only make a small contribution to the final estimate. 
The average work $\overline{W} =  \Delta F + \overline{W}_d$, in which $\overline{W}_d$ is the average dissipated work; the probability of carrying out a dominant realization, for which $W \approx \Delta F$, is much lower.
Indeed, the free energy estimator can be very difficult to converge in the context of molecular dynamics simulations, with unpredictable bias and variance~\cite{Hummer2001,Gore2003,Park2004,Shirts2005,Lua2005,Oberhofer2005,Jarzynski2006}.
The large number nonequilibrium trajectories, and associated computational cost, required for a reliable application of Eq.~\eqref{eq:Jarzynski} makes its usefulness for free energy calculation impractical.
Therefore, alternative strategies have been proposed that increase the probability of sampling dominant, low-work trajectories or use more robust free energy estimators~\cite{Ytreberg2004,Echeverria2012,Wolf2018,Arrar2019,Roussey2020}.

Within MD simulations, the transformation $\chi_a \rightarrow \chi_b$ can be realized using steered MD (SMD), by coupling a moving Hooke spring potential to $\chi$.
Park and Schulten observed that in the limit of stiff springs and overdamped Langevin dynamics, $W$ follows a Gaussian distribution, centered at $\overline{W}$~\cite{Park2004}.
In this regime, the second order cumulant expansion therefore becomes a good estimator of $\Delta F$:
\begin{equation}
  F(\chi) \approx \langle W(\chi) \rangle - \frac{\beta}{2} \left (\langle W(\chi)^2 \rangle - \langle W(\chi) \rangle^2 \right) , \label{eq:cumulant}
\end{equation}
where we have chosen $F(\chi_a) = 0$.
In principle, this expression exhibits better convergence when only a small sample of SMD trajectories is available, although some reports have doubted its practical accuracy~\cite{Noh2020}.

\subsection{A reweighting approach}

Here, we propose to sidestep the convergence concerns regarding the application of the Jarzynski equality to SMD simulations.
We do not attempt to directly calculate the free energy $F$ from a sample $\{W_i\}$.
Rather, we use the Jarzynski equality to learn a bias potential $V(\chi)$, and reconstruct a free energy profile $F(\chi)$ from a reweighted molecular dynamics trajectory on the biased potential energy surface. 
To wit, our procedure is as follows:
\begin{enumerate}
    \item Perform a small number of SMD simulations, pulling the system from $\chi_a$ to $\chi_b$. 
    For each trajectory $i$, record the instantaneous values of $(\chi_i(t_j), W_i(t_j))$ at regular time intervals $t_j$.
    \item Learn smooth functions $W_i(\chi)$ from the respective set of $(\chi_i, W_i)$ data. 
    Use the cumulant expression Eq.~\eqref{eq:cumulant} to calculate an estimated free energy profile $\widetilde{F}(\chi)$ from $\{ W_i(\chi) \}$.
    \item Perform a MD simulation while applying an external bias potential $V(\chi) = -\widetilde{F}(\chi)$. 
    Reconstruct the true free energy surface $F(\chi)$ from the reweighted histogram $\langle w(t) \cdot \delta [\chi - \chi(t)] \rangle_b$, in which $w(t) = e^{\beta V(\chi(t))}$ and $\langle \cdots \rangle_b$ denotes a time average on the biased potential energy surface.
\end{enumerate}

In order to maximize the validity of Eq.~\eqref{eq:cumulant} all SMD simulations should be carried out with a strongly coupled Langevin thermostat and use spring forces that can tightly confine $\chi$ throughout the trajectory.
As a rule of thumb, the aim was to limit fluctuations of $\chi$ to around 1\% of $|\chi_b - \chi_a|$ in all reported simulations.
Still, the resulting $(\chi_i, W_i)$ will be noisy.
Dealing with such data is a key aspect of machine learning techniques, and we use kernel ridge regression to learn smooth functions $W_i(\chi)$ from each SMD run $i$.
We can then obtain an approximate free energy surface $\widetilde{F}(\chi)$ by inserting all $W_i(\chi)$ in the cumulant expansion of Eq.~\eqref{eq:cumulant}.

We do not need $\widetilde{F}(\chi)$ to be a high-quality estimator of $F(\chi)$.
The bias potential $V(\chi) = -\widetilde{F}(\chi)$ will sufficiently enhance sampling in the interval $[\chi_a, \chi_b]$ within MD time scales as long as it approximates the true free energy up to a few $k_B T$.
The curvature of $\widetilde{F}(\chi)$ is strongly dependent on the process, which can pose a challenge when using it to calculate a biasing force.
Here, too, machine learning techniques can provide an outcome: Artificial neural networks (ANNs) can efficiently represent arbitrary free energy surfaces in a compact and differentiable manner~\cite{Schneider2017,Galvelis2017,Sidky2018,Bonati2019}.
We find that the one-dimensional $\widetilde{F}(\chi)$ (or $V(\chi)$) is always well-described by an ANN with a single hidden layer.
Note that the use of an ANN is not a crucial aspect of our approach, but is merely the most convenient option available to us that avoids the discretization and boundary errors associated with storing $\widetilde{F}$ on a grid.

$F (\chi)$ is defined in terms of the marginal probability density of $\chi$, $p(\chi) =\langle \delta [\chi - \chi(t)] \rangle$:
\begin{equation}
  F (\chi) = -\frac{1}{\beta} \ln \langle \delta [\chi - \chi(t)] \rangle .
\end{equation}
Accordingly, if the external bias potential $V (\chi)$ is used to enhance sampling along $\chi$, reweighting of the biased trajectory recovers $p(\chi)$~\cite{Torrie1977}:
\begin{equation}
  F (\chi) = -\frac{1}{\beta} \ln \langle w(t) \cdot \delta [\chi - \chi(t)] \rangle_b , \label{eq:reweight}
\end{equation}
in which $w(t) = e^{\beta V(\chi(t))}$ and $V(\chi) = -\widetilde{F}(\chi)$, as noted before.
Because the bias potential $V (\chi)$ is static in our case, the simple reweighting formula above suffices.
Alternatively, $F$ can be recovered directly from the biased histogram $H(\chi) = \langle \delta [\chi - \chi(t)] \rangle_b$ through
\begin{equation}
  F (\chi) = \widetilde{F}(\chi) - \frac{1}{\beta} \ln \langle \delta [\chi - \chi(t)] \rangle_b , \label{eq:altreweight}
\end{equation}
which might exhibit better numerical stability if $\widetilde{F}$ is large.
The form of eq.~\eqref{eq:altreweight} also highlights that the biased sampling run serves to correct the imperfect nature of $\widetilde{F}(\chi)$: If $\widetilde{F}(\chi) = F(\chi)$, $V(\chi) = - \widetilde{F}(\chi)$ is a perfect bias potential and $H(\chi)$ will be flat.

We have restricted ourselves to a one-dimensional FES $F(\chi)$.
For many systems and transformations, good approximate reaction coordinates $\chi$ are known and our method can be readily applied.
However, even if no intuitive reaction coordinate is available, recent data-driven approaches have made it possible to discover candidate reaction coordinates $\chi (\mathbf{R})$ in an automated manner~\cite{Sidky2020}.
In one following example (sec.~\ref{sec:HBr}), we show how one such method---harmonic linear discriminant analysis (HLDA)~\cite{Mendels2018}---can be integrated in our workflow.
 
\section{Examples}

\subsection{General computational details}

All SMD simulations, sampling, and reweighting were carried out using PLUMED~\cite{Tribello2014,PLUMED2019} interfaced with different simulation codes: CP2K~\cite{Kuhne2020} in case of the S$_\mathrm{N}$2 and hydrobromination reactions and LAMMPS~\cite{Plimpton1995} for the nucleation process and solvated dimer system.
The temperature was controlled via Langevin thermostats: a white noise implementation~\cite{Bussi2007} in the LAMMPS-based simulations, and a colored noise generalized Langevin thermostat~\cite{Ceriotti2009} in CP2K, optimized for efficient sampling~\cite{Ceriotti2010}.

All machine learning algorithms were used as implemented in the scikit-learn library~\cite{Pedregosa2011}.
Each $W_i (\chi)$ was interpolated with kernel ridge regression, using a regularization strength of $0.1$ and a radial basis function (RBF) kernel; the learned models were then used to generate a large number $W_i(\chi)$ values in the interval $[\chi_a,\chi_b]$ (in the order of $10^3$ to $10^4$, depending on the system).
From these data points, the cumulant estimator of $\widetilde{F} (\chi)$ was calculated at all considered values of $\chi$.
A single hidden layer ANN with $\tanh$ activation functions was then fitted to these $(\chi, \widetilde{F}(\chi))$ value pairs.
This training data was shuffled prior to learning the ANN, and $10$\% of the data points were held back for validation purposes.
The number of neurons in the hidden layer was chosen to minimize the error on the test set.

For each of the considered systems, only five independent SMD trajectories were computed.
Three independent reweighting simulations were carried out for each system to assess their reproducibility.
Error bars on free energy differences or barriers are then reported at the 75~\% confidence level.

Free energy differences $\Delta F$ between metastable states $A$ and $B$ are defined as:
\begin{equation}
  \Delta F = F_B - F_A \quad \text{with} \quad F_X = -\frac{1}{\beta} \int_X \mathrm{d} \chi \, e^{-\beta F(\chi)} ,
\end{equation}
whereas the free energy barrier of a transition $A \rightarrow B$, passing through a dividing surface $\chi = \chi^{*}$, is
\begin{equation}
  \Delta^\ddagger F = F(\chi^*) + \frac{1}{\beta} \ln \frac{\langle |\nabla \chi| \rangle^{-1}_{\chi = \chi^*}}{h} \sqrt{\frac{2 \pi m}{\beta}} - F_A , \label{eq:barrier}
\end{equation}
with $h$ being the Planck constant and $m$ the mass of the reaction coordinate~\cite{Bal2020JCP}.
The rate constant of the process can then be calculated within the framework of transition state theory (TST), from the Eyring expression $k^\mathrm{TST} = (h \beta)^{-1} e^{-\beta \Delta^\ddagger F}$.
The definition of the barrier in Eq.~\eqref{eq:barrier} guarantees that the choice of the dividing surface $\chi = \chi^*$ that maximizes $\Delta^\ddagger F$ also minimizes the TST rate $k^\mathrm{TST}$, which is in turn an upper bound on the true rate $k$.
A gauge correction, function of the norm of the gradient of $\chi$ with respect to all atomic coordinates $\mathbf{R}$, is needed to obtain barriers invariant to the functional form of $\chi (\mathbf{R})$.

Sample inputs and scripts to reproduce the reported simulations are deposited on PLUMED-NEST, the public repository of the PLUMED consortium~\cite{PLUMED2019}, as plumID:21.020~\cite{data}.

\subsection{Symmetric nucleophilic substitution reaction}

\begin{figure*}[t]
\includegraphics{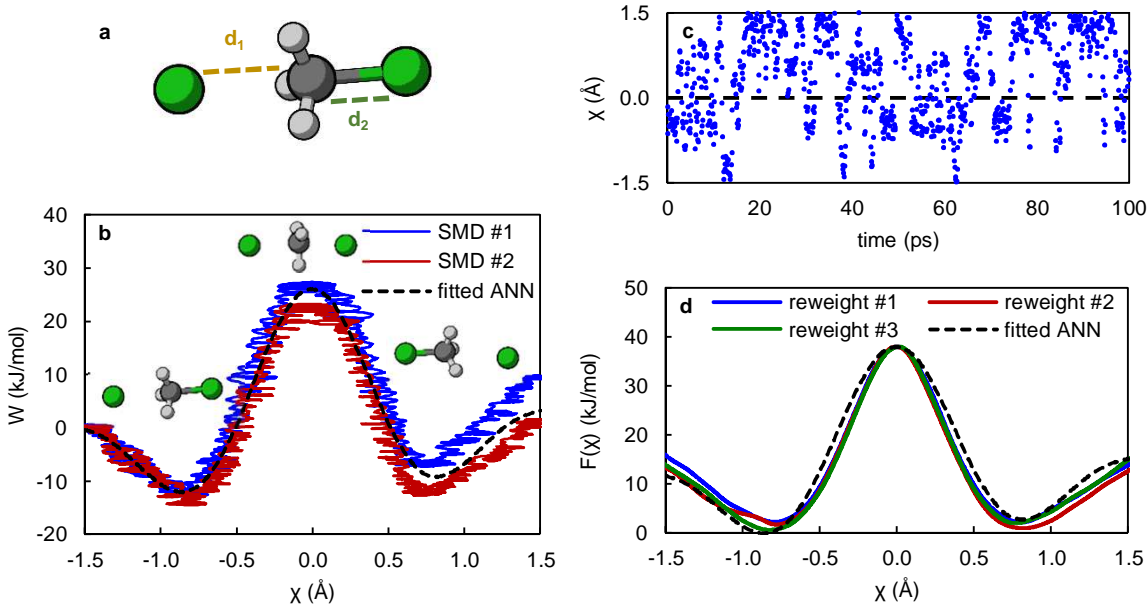}
\caption{\label{fig:panel-sn2} Symmetric S$_\mathrm{N}2$ reaction of methyl chloride.
(a) Definition of the distances $d_1$ and $d_2$.
(b) $(\chi,W)$ of two independent SMD runs, and the final ANN representation of $\widetilde{F} (\chi)$.
(c) Example trajectory of $\chi$ during a reweighting run using $-\widetilde{F} (\chi)$ as external bias potential.
(d) Final estimates $F (\chi)$ from three independent reweighting runs, compared to $\widetilde{F} (\chi)$.
The curves are vertically aligned at $\chi = 0$~\AA{}.}
\end{figure*}

As a first illustration of reweighted Jarzynski sampling, we use the symmetric S$_\mathrm{N}2$ reaction of methyl chloride, $\mathrm{Cl^-} + \mathrm{CH_3Cl} \rightarrow \mathrm{ClCH_3} + \mathrm{Cl^-}$, a common testbed for enhanced sampling techniques~\cite{Ensing2005,Fleming2016,Fu2017,Zhang2019,Bal2020JPCL}.
Intuitively, we choose $\chi = (d_2 - d_1)/\sqrt{2}$ (Figure~\ref{fig:panel-sn2}a).
We apply harmonic restraints to keep $d_1$ and $d_2$ below 5~\AA{}, describe the interatomic interactions with the semi-empirical PM6 Hamiltonian~\cite{Stewart2007}, and integrate the equations of motion with a time step of 0.5~fs.
SMD simulations were then performed in the interval $\chi \in [-1.5, 1.5]$~\AA{} at $T = 300$~K, over a time scale of 10~ps.

The $(\chi (t), W(t))$ trajectory of two independent SMD simulations can be quite different (up to about $2 k_B T$), as shown in Figure~\ref{fig:panel-sn2}b.
Even though the sampled data is rather noisy (using a spring constant of $10^3$~kJ~mol$^{-1}$~\AA{}$^{-2}$), the final ANN approximation of $\widetilde{F} (\chi)$ is however a smooth curve, for which 12 hidden neurons suffice.
Because of the symmetry of the problem, we can also see that $\widetilde{F} (\chi)$ is not a perfect estimator of the true FES, as it is biased towards the $\chi < 0$ state by about 3~kJ/mol.
Such a bias is usually hard to distinguish from the variance or other sources of error~\cite{Gore2003,Shirts2005}.
The asymmetry of the putative FES also affects a reweighting run using $V (\chi) = -\widetilde{F} (\chi)$ as a bias potential, which preferentially samples the $\chi > 0$ state (Figure~\ref{fig:panel-sn2}c).
The bias potential does, however, also efficiently induce many state-to-state crossings, and an extensive sampling of the transition state region at $\chi \approx 0$.
As a result, the final estimates of $F (\chi)$ after 100~ps reweighting runs are, on average, free from the initial bias in $\widetilde{F} (\chi)$ (Figure~\ref{fig:panel-sn2}d): From three trajectories, we find $\Delta F = 0.2 \pm 1.1$~kJ/mol.

A very modest sampling time of $5 \times 10$~ps (SMD runs) + 100~ps (reweighting) can thus yield reaction free energies with rather good precision, and repeated reweighting runs show that the uncertainty is well below $k_B T$.
This does not only hold true for thermodynamics, but also kinetics.
The free energy barrier of the forward reaction (from low to high $\chi$) is predicted to be $\Delta^\ddagger F = 36.7 \pm 0.8$~kJ/mol, corresponding to a rate $k^\mathrm{TST} = (2.6 \pm 0.8) \times 10^6$~s$^{-1}$.
This implies an average reaction time of $390 \pm 120$~ns, which compares well to the range 180--300~ns obtained by Fu et al. using metadynamics in an almost identical model system~\cite{Fu2017}.
TST thus appears to describe this particular process rather well.

The ability of the reweighting procedure to quickly and transparently remove the bias error that may be present in $\widetilde{F}$ is greatly facilitated by the static nature of the bias.
Whenever many state-to-state transitions are observed during the reweighting run, in principle the only remaining error in $F$ will be due to variance.
The size of this variance can then be estimated with the usual techniques, such as block averaging or repeated simulations.

\subsection{Droplet nucleation in supersaturated vapor}

\begin{figure}[tb]
\includegraphics{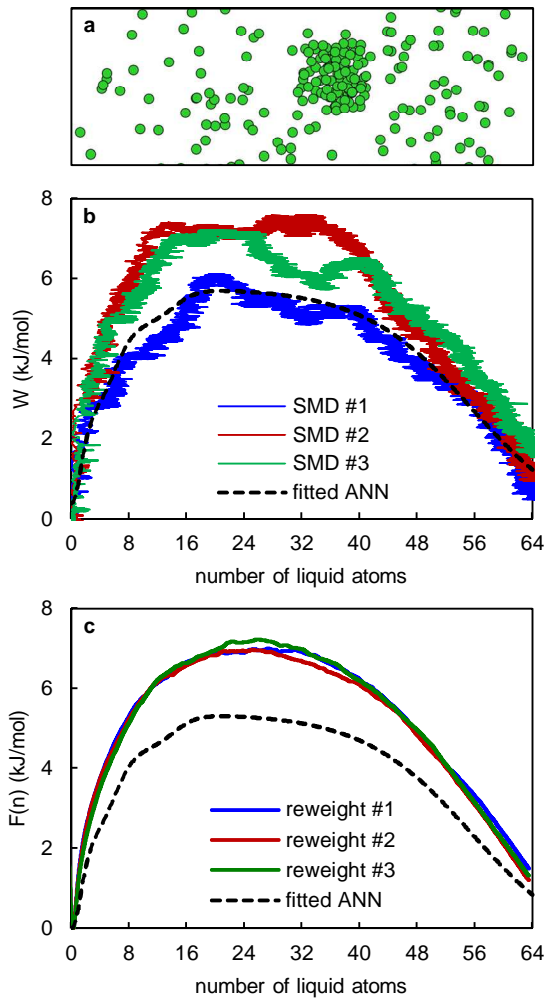}
\caption{\label{fig:panel-droplet} Droplet nucleation from supersaturated Ar vapor.
(a) Snapshot of a configuration corresponding to $n \approx 64$.
(b) Example $(n,W)$ curves of three independent SMD runs, and the final ANN representation of $\widetilde{F} (n)$.
(c) Final estimates $F (n)$ from three independent reweighting runs, compared to $\widetilde{F} (n)$.
The curves are vertically aligned at $n=0$.}
\end{figure}

As an example of a process involving diffusive collective motion of a large number of atoms, we revisit the condensation of Ar vapor at $T=80.7$~K, for which rates have been calculated previously~\cite{Salvalaglio2016,Bal2021}.
We model 512 atoms in a cubic simulation cell of length 11.5~nm, corresponding to an initial supersaturation of 8.68.
Interatomic interaction are described by a Lennard-Jones potential, using the same parameters as earlier work~\cite{Salvalaglio2016}.
We use the number of liquidlike atoms $n$ as a reaction coordinate, that is, the number of atoms that have more than five nearest neighbors.
Equations of motion are integrated with a time step of 5~fs.
Because we are only interested in the formation of the critical nucleus, we perform SMD from $n = 0$ to $n = 64$ over 10~ns, and also place a harmonic wall at $n = 64$ (Figure~\ref{fig:panel-droplet}a).

The individual $(n(t), W(t))$ trajectories can differ quite strongly for the process, and are also rather rough (Figure~\ref{fig:panel-droplet}b).
$\widetilde{F} (n)$ is however again a very smooth function owing to the regularization employed in the kernel ridge regression interpolation, and is well represented by only 12 neurons in the hidden layer.
From three 100~ns reweighting runs we can see that: (1) $\widetilde{F} (n)$ mostly recovers the shape of the FES, and while it underestimates the  nucleation barrier by  more than 2~$k_B T$ (2) the reweighted $F (n)$ reliably converges within this time scale in a very consistent manner (Figure~\ref{fig:panel-droplet}c).

Nucleation barriers are recovered with very high precision, yielding the estimate $\Delta ^\ddagger F = 7.06 \pm 0.15$~kJ/mol.
Note that in our previous study of this process~\cite{Bal2021}, a metadynamics simulation required 1~$\mu$s to produce a barrier estimate with a larger uncertainty, namely, $7.46 \pm 0.43$~kJ/mol (error calculated from four 200~ns trajectory chunks, adjusted to the same confidence level of 75~\%).
A correct rate estimate also requires knowledge of the transmission coefficient, which is $\kappa = (2.1 \pm 0.5) \times 10^{-3}$ at this particular supersaturation level~\cite{Bal2021}.
Any rate estimate $k = \kappa k^\mathrm{TST}$ is only valid within the simulation cell volume $V$.
A global nucleation rate should therefore be calculated as $J = k/V$~\cite{Salvalaglio2016,Bal2021}.
From this data, we can estimate the nucleation rate to be $J = (6.3 \pm 2.1) \times 10^{22}$~cm$^{-3}~s^{-1}$, close to our previously reported value of $(3.5 \pm 2.4) \times 10^{22}$~cm$^{-3}~s^{-1}$.
The overall computational cost required in this work is however three times lower.

\subsection{Dissociation and association in solution}

Processes in solution depend on an intricate interplay between solvent and solute, which makes their analysis non-trivial~\cite{Mullen2014}.
We introduced a simple example of such a system in earlier work, consisting of a Morse dimer solvated in a dense Lennard--Jones fluid~\cite{Bal2020JCP}.

\begin{figure}[tb]
\includegraphics{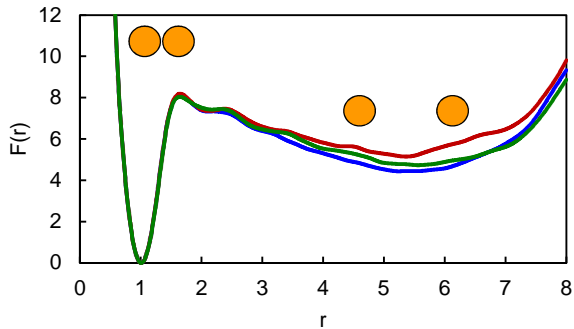}
\caption{\label{fig:panel-dimer} Free energy of a strongly interacting dimer, solvated by a Lennard-Jones fluid, as a function of the interatomic separation $r$.
The resultant $F(r)$ of three independent reweighting runs are depicted in different colors.}
\end{figure}

Reduced units were used throughout.
We performed SMD simulations on this system at a reduced temperature of $T = 1.5$, manipulating the interatomic distance in the dimer from $r = 0.8$ to $r = 8.0$, over a time scale of $10^3 \tau$ ($5 \times 10^5$ time steps).
We again fit $\widetilde{F} (r)$ to a \{12\} ANN and converge $F (r)$ in reweighting runs of $10^4 \tau$, depicted in Figure~\ref{fig:panel-dimer}.

$F (r)$ is highly asymmetric, and it seems that narrow deep free energy wells (i.e., the bound state) are better-sampled than wide, shallow states (the dissociated state).
This is also reflected by the unbinding free energy barrier $\Delta^\ddagger F = 4.69 \pm 0.07$, which is determined with a higher precision than the unbinding free energy, $\Delta F = 1.17 \pm 0.33$.
These estimates are in agreement with previously calculated values, which are $\Delta^\ddagger F = 4.83 \pm 0.14$ and $\Delta F = 1.39 \pm 0.16$, respectively~\cite{Bal2020JCP}.

\subsection{Four-center addition reaction}
\label{sec:HBr}

It is not always trivial to find an appropriate one-dimensional reaction coordinate $\chi$.
For example, the hydrobromination reaction of propene, HBr~+ C$_3$H$_6$, involves forming a C--Br bond, forming a C--H bond, breaking the H--Br bond, and changing the bonding character of the vinyl moiety.
In addition, two possible products can be formed: the dominant 2-bromopropane (the Markovnikov product) and the less likely 1-bromopropane (the anti-Markovnikov product).
Due to this manybody nature, the reaction has recently been recognized as an interesting test case for targeted sampling of reaction pathways.
Methods that aim to explore high-dimensional configuration spaces have been tested on this system~\cite{Debnath2020}, as well as dimensionality reduction techniques~\cite{Piccini2018}.

\begin{figure*}[t]
\includegraphics{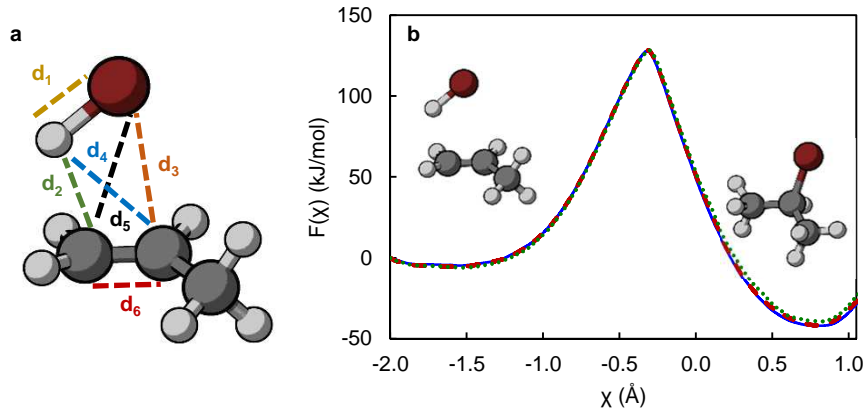}
\caption{\label{fig:panel-hbr} Hydrobromination of propene.
(a) Definition of the six degrees of freedom included in the reaction coordinate $\chi (\mathbf{R})$.
(b) Free energy surface $F (\chi)$ of the formation of the Markovnikov product 2-bromopropane.
The resultant $F (\chi)$ of three independent reweighting runs are depicted in different colors and line styles.}
\end{figure*}

Here, we wish to specifically target the Markovnikov pathway.
For this purpose, we must identify a suitable reaction coordinate $\chi (\mathbf{R})$.
A method that integrates rather well in our established workflow, is harmonic linear discriminant analysis (HLDA)~\cite{Mendels2018}.
In HLDA, one first identifies the metastable states of interest, which are here the HBr~+ propene system and 2-bromopropane, respectively.
A set of simple microscopic descriptors that may be able to discriminate between these states must then be selected.
In order to exploit chemical intuition as little as possible, we here opt to use all internal degrees of freedom of the four reaction centers, namely, the six interatomic distances between H, Br and the vinylic C atoms (Figure~\ref{fig:panel-hbr}a).
Finally, short MD runs are carried out in the two metastable state of interest so that for each state $i$, the mean $\boldsymbol{\mu}_i$ and multivariate variance $\boldsymbol{\Sigma}_i$ of the descriptors $\mathbf{d} = (d_1,d_2,\ldots,d_6)$ can be calculated.
Finally, the coordinate $\chi$ that best separates the two states $A$ and $B$ is calculated as a linear combination of the candidate descriptors:
\begin{equation}
  \chi = (\boldsymbol{\mu}_A - \boldsymbol{\mu}_B)^T \left ( \frac{1}{\boldsymbol{\Sigma}_A} + \frac{1}{\boldsymbol{\Sigma}_B} \right ) \mathbf{d} ,
\end{equation}
in a modification of Fisher's linear discriminant analysis (LDA).
An advantage of HLDA is that it only requires us to know \textit{a priori} the metastable states of interest, without having to sample any $A \rightarrow B$ transition pathway first.

As in previous studies~\cite{Piccini2018,Debnath2020}, harmonic restraints were used to keep all components of $\mathbf{d}$ below 3~\AA{}, interatomic forces were calculated at the semi-empirical PM6 level of theory~\cite{Stewart2007}, and sampling was carried out at $T = 300$~K.
The integration time step was 0.5~fs.
We first parameterized $\chi = 0.661 d_1 - 0.656 d_2 - 0.328 d_3 + 0.011 d_4 - 0.021 d_5 + 0.157 d_6$ from 20~ps sampling runs in the metastable states.
Then, SMD runs were carried out over 10~ps from $\chi = -2$~\AA{} and $\chi = 1$~\AA{}.
A larger \{48\} ANN was found to be needed for an accurate representation of the sharply varying $\widetilde{F} (\chi)$.
Finally, reweighting over just 100~ps yields very consistent estimates of the FES $F (\chi)$ (Figure~\ref{fig:panel-hbr}b).

The predicted reaction free energy $\Delta F = -33.8 \pm 1.9$~kJ/mol, and the barrier $\Delta^\ddagger F = 131.7 \pm 0.8$~kJ/mol.
This very high free energy barrier implies that the reaction is very slow: $k^\mathrm{TST} = (7.46 \pm 2.34) \times 10^{-11}$~s$^{-1}$.
However, the bias potential $-\widetilde{F} (\chi)$ is of sufficient quality to induce several transitions in both directions within short MD runs.

Due to its high barriers and asymmetric FES this particular system constitutes the most challenging application of reweighted Jarzynski sampling up till now.
It is therefore an instructive test case for comparing the efficiency of our sampling recipe to state-of-the-art adaptive techniques.
For this purpose, we choose metadynamics,~\cite{Laio2002} variationally enhanced sampling (VES),~\cite{Valsson2014} and the on-the-fly probability-enhanced sampling (OPES) method~\cite{Invernizzi2020}.
A fair comparison is easiest to make with OPES due to its limited number of parameters: Only the bias update frequency and an estimate of the highest barrier is required for it to be applied.
We can stack the deck strongly in the favor of OPES because we already have an estimate of the barriers by now: We set the OPES barrier parameter to 200~kJ/mol, well above either barrier of interest, and update the bias in 100~fs intervals.
Due to their larger number of user-set parameters, a straightforward comparison to metadynamics and VES is somewhat more tricky.
We opt to use a rather aggressive biasing strategy (along the lines of previous metadynamics simulations on the system~\cite{Piccini2018}): A hill height/bias update stepsize of 5~kJ/mol, a well-tempered bias factor~\cite{Barducci2008,Valsson2015} of $\gamma = 50$, and bias update stride of 100~fs.

\begin{figure*}[t]
\includegraphics{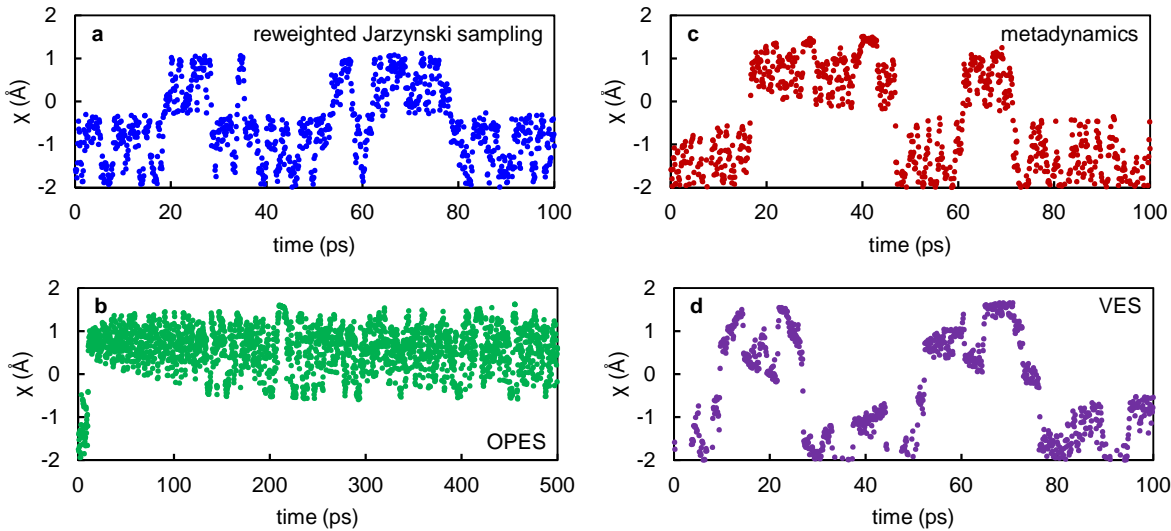}
\caption{\label{fig:panel-opes} Comparing sampling efficiency of different methods in the hydrobromination reaction.
Trajectory of $\chi$ in (a) a reweighting run using $-\widetilde{F} (\chi)$ as static bias and adaptive runs with (b) OPES, (c) metadynamics, and (d) VES.
Note that the depicted OPES trajectory is five times longer.}
\end{figure*}

We plot the trajectory of $\chi$ in a reweighting run under influence of the bias $-\widetilde{F} (\chi)$ as well as during the various dynamic biasing runs (Figure~\ref{fig:panel-opes}).
As noted earlier, the Jarzynski-derived $\widetilde{F} (\chi)$ is a good model for both barriers and leads to nearly uniform sampling from the get-go (Figure~\ref{fig:panel-opes}a).
This is not true for OPES (Figure~\ref{fig:panel-opes}b).
The initial $A \rightarrow B$ transition occurs rather quickly in the OPES run, because its barrier height is approximately known and an efficient bias can be easily constructed.
Then, however, the simulation remains stuck in the $B$ state: The OPES algorithm has no way of knowing the free energy difference $\Delta F$ between the metastable states or, equivalently, the barrier height of the $B \rightarrow A$ transition.
The bias for escaping state $B$ is therefore only accumulated slowly.
Similar behaviour is also observed with slower bias update frequencies (1~ps) and a lower barrier parameter (150~kJ/mol).

\begin{table}[tb]
  \caption{Comparison of Hydrobromination Free Energy Estimates from Different Sampling Methods\textsuperscript{\emph{a}}}
  \label{tab:hbr-comp}
  \begin{tabular}{lccc}
  \hline
  Method            & $t$ (ps) & $\Delta F$  (kcal/mol) & $\Delta^\ddagger F$ (kcal/mol) \\
  \hline
  This work    & 100           & $-33.8 \pm 1.9$        & $131.7 \pm 0.8$ \\
  Metadynamics & 100           & $4.0 \pm 0.7$          & $153.9 \pm 3.8$ \\
               & 500           & $-22 \pm 14$           & $134.1 \pm 1.8$ \\
  VES          & 100           & $-56.8 \pm 4.9$        & $124 \pm 16$ \\\
               & 500           & $-37 \pm 13$           & $125.7 \pm 6.6$ \\
  \hline
  \end{tabular}

  \textsuperscript{\emph{a}} Reaction free energy $\Delta F$ and barrier $\Delta^\ddagger F$ from reweighting runs of varying length $t$.
\end{table}

Metadynamics and VES fare better (Figs.~\ref{fig:panel-opes}c--d). 
Both algorithms can reliably drive transitions within 100~ps, which is about on par with the Jarzynski bias.
We can therefore also compare the quality of their respective free energy estimates.
For fair comparison we use reweighting techniques~\cite{Tiwary2015,Valsson2014} to reconstruct the free energy estimates also for these methods.
Three independent simulations are performed with each algorithm.
Reweighting the action of a fluctuating bias potential is more difficult~\cite{Cuendet2014,Mones2016,Marinova2019}, as can be seen in Table~\ref{tab:hbr-comp}.
A 100~ps sampling time is not sufficient to produce quantitatively correct free energy estimates.
Longer 500~ps sampling runs improve the quality of $\Delta F$ and $\Delta^\ddagger F$, but still with a substantially higher variance than reweighted Jarzynski sampling; better reweighting approaches or biasing strategies might however improve the performance of metadynamics and VES.

\subsection{Limitations and future prospects}

Although the good performance of reweighted Jarzynski sampling in the above examples is very encouraging, we must address two obvious limitations of the approach.

First, the approach outlined here is intended for the targeted calculation of free energy differences and barriers of transformations along a single reaction coordinate.
The reaction path of interest must be known \textit{a priori}.
The study of high-dimensional free energy surfaces is therefore not the main application domain.
Nor is the unconstrained exploration of configuration space:
To characterize previously unknown states and pathways, one can use specialized ``blind'' methods~\cite{Bal2016,Fu2018,Debnath2020,Giberti2021}.

\begin{figure*}[t]
\includegraphics{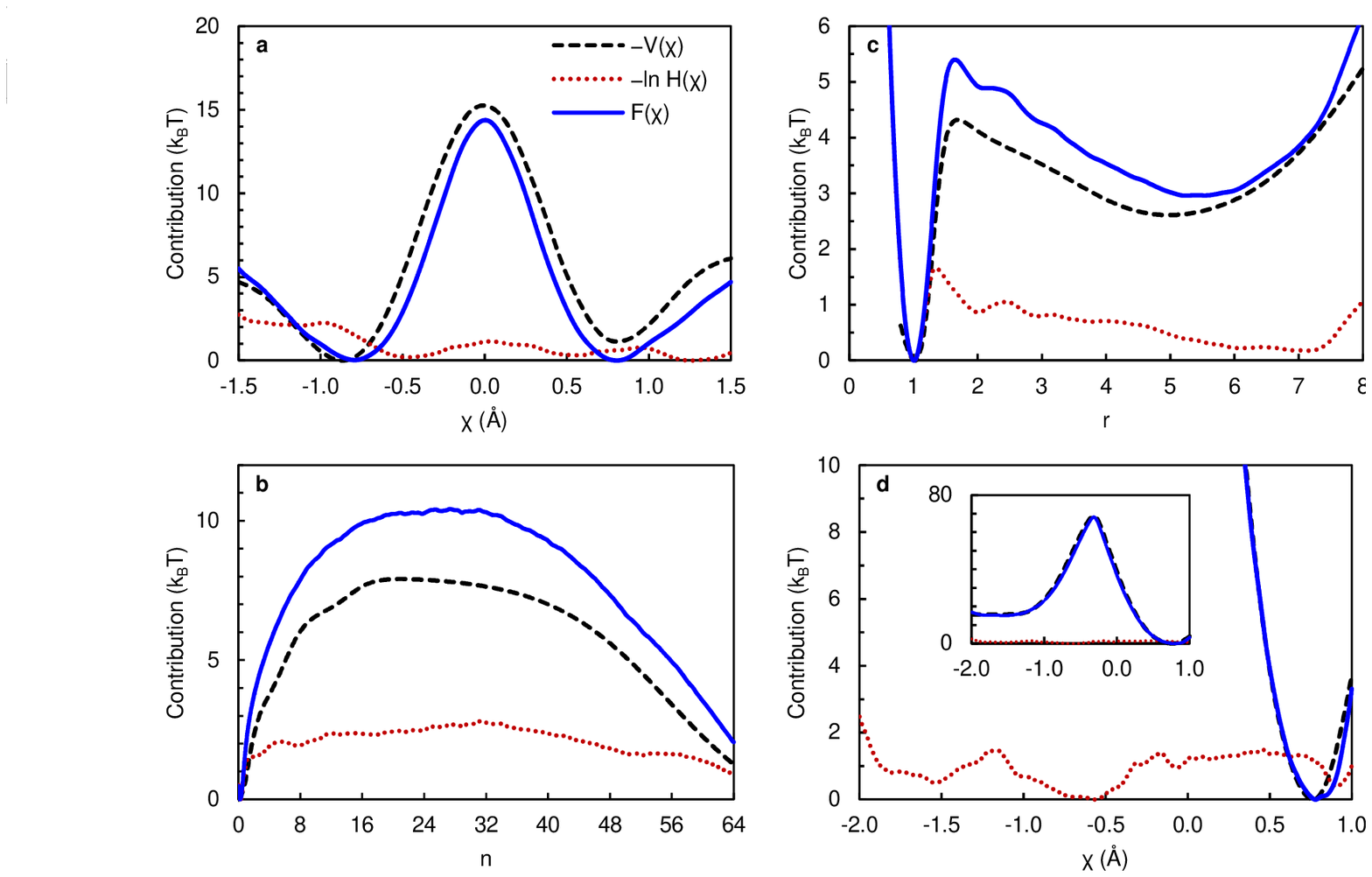}
\caption{\label{fig:panel-efficiency} The ANN $\widetilde{F}$ approximates $F$ up to a few $k_B T$ in the considered systems.
Using the the bias $V = -\widetilde{F}$, the FES is almost fully flattened, as demonstrated by the biased histogram $H = \langle \delta [\chi - \chi(t)] \rangle_b$.
The contributions to the FES estimate $F (\chi) = -V(\chi) - k_B T \ln H (\chi)$ are shown in units of $k_B T$ for (a) the symmetric S$_\mathrm{N}$2 reaction, (b) droplet nucleation, (c) the solvated dimer, and (d) the hydromination reaction.}
\end{figure*}

Second, although $-\widetilde{F}$ has turned out to be a perfectly cromulent bias potential in the above examples, this might not be generally true.
As noted earlier, it is difficult to consistently eliminate bias and variance errors in applications of the Jarzynski equality~\cite{Gore2003,Shirts2005}.
If $\widetilde{F}$ deviates from $F$ by more than a few $k_B T$, the resultant bias may not lead to sufficient sampling.
The above examples therefore represent more or less ideal scenarios: the ANN $\widetilde{F}(\chi)$ is a good model of $F(\chi)$, and the biased histogram $H (\chi) = \langle \delta [\chi - \chi(t)] \rangle_b$ is nearly flat (Figure~\ref{fig:panel-efficiency}).
The correction $- k_B T \ln H (\chi)$ to $\widetilde{F}(\chi)$ (following Eq.~\eqref{eq:altreweight}) is therefore always only a few $k_B T$.
Because it is however not \textit{a priori} obvious what an optimal SMD set up is for a specific system, such a high-quality $\widetilde{F}$ may not possible to obtain reliably for all processes.
One possible way to make $-\widetilde{F}$ an even more robust bias would be to use methods with improved convergence~\cite{Ytreberg2004,Echeverria2012,Wolf2018,Arrar2019,Roussey2020}.
In addition, the toolbox of nonequilibrium nanoscale thermodynamics~\cite{Jarzynski2011} could be further explored:
For example, if also reverse trajectories ($B \rightarrow A$) are sampled, the Crooks fluctuation theorem~\cite{Crooks1999} can be invoked as an alternative means of estimating free energies.

Even if $\widetilde{F}$ is a poor approximation of $F$ it can still serve as an initial step in an enhanced sampling procedure:
An adaptive sampling method like metadynamics, variationally enhanced sampling, or OPES could operate on top of a preconditioned, mollified FES $F - \widetilde{F}$.
Such preconditioning may also help prevent the adaptive method from getting stuck in very deep minima.
Alternatively, following other recent strategies, our approach could be the basis of an iterative procedure, in which the initial estimate $\widetilde{F}$ is further refined in a sequence of sampling runs until convergence is reached~\cite{Galvelis2017,Sidky2018,Invernizzi2020}---a strategy that dates back to at least adaptive umbrella sampling~\cite{Mezei1987}.

Currently, we have only employed a \emph{parallel} strategy to improve the estimate of $F$, in the sense that multiple independent reweighting runs were carried out.
If we observed that several transitions occurred within each reweighting run, we assumed that the bias error in $F$ was small, so that we could directly quantify the inherent variance of the FES estimates by comparing the reweighting simulations.
If either check proved unsatisfactory for our trial runs, we simply opted to carry out the reweighting runs for more steps.
We also did not explicitly asses the quality of the employed reaction coordinates: A poorly chosen $\chi$ negatively affects the performance of adaptive sampling methods, and this is equally true for our approach.

Finally, we would like to note one very practical convenience of the reweighted Jarzynski sampling approach as described here:
All required simulation steps can be readily carried out with unmodified versions of standard sampling codes.
The only requirement is that moving harmonic restraints and the collective variables of choice are implemented, and that arbitrary bias potentials are allowed.
Recent versions of PLUMED~\cite{Tribello2014,PLUMED2019} possess these features.

\section{Conclusions}

We have introduced a recipe for the calculation of a one-dimensional free energy surface (FES) from atomistic simulations.
The final FES estimate $F (\chi)$ is obtained from the reweighted \emph{equilibrium} probability density of the coordinate $\chi$, sampled under influence of an external bias potential $V (\chi)$.
The bias $V (\chi)$, in turn, is learned first from a small collection of \emph{nonequilibrium} trajectories by employing the Jarzynski equality.
$V (\chi)$ is smoothly and compactly represented in the form of an artificial neural network (ANN).

We find reweighted Jarzynski sampling to exhibit excellent efficiency and accuracy in the considered systems.
Free energy differences and barriers with error bars well below $k_B T$ are routinely obtained within very economical simulation time scales, and accurate kinetics are simultaneously accessible through a recent numerical framework~\cite{Bal2020JCP,Bal2021}.

The approach is also compatible with adaptively biased sampling methods so that a joint application may exhibit faster convergence.
The method can readily use any type of collective variable as reaction coordinate $\chi$, and only requires an off-the-shelf version of the PLUMED plugin or any other free energy code with similar capabilities.

\section*{Acknowledgments}
K.M.B. was funded as a junior postdoctoral fellow of the FWO (Research Foundation -- Flanders), Grant 12ZI420N.
The computational resources and services used in this work were provided by the HPC core facility CalcUA of the Universiteit Antwerpen, and VSC (Flemish Supercomputer Center), funded by the FWO and the Flemish Government.
HLDA calculations were performed with a script provided by GiovanniMaria Piccini.
K.M.B. thanks Erik Neyts for proofreading this manuscript, the many discussions over the years, and for his continuous support.

\bibliography{bibliography}

\end{document}